\documentclass{article}

\usepackage{arxiv}

\usepackage[utf8]{inputenc} % allow utf-8 input
\usepackage[T1]{fontenc}    % use 8-bit T1 fonts
\usepackage{hyperref}       % hyperlinks
\usepackage{url}            % simple URL typesetting
\usepackage{booktabs}       % professional-quality tables
\usepackage{amsfonts}       % blackboard math symbols
\usepackage{nicefrac}       % compact symbols for 1/2, etc.
\usepackage{microtype}      % microtypography
\usepackage{lipsum}		% Can be removed after putting your text content
\usepackage{graphicx}
\usepackage{natbib}
\usepackage{doi}

\title{CustOmics: A  versatile deep-learning based strategy for multi-omics integration}

\author{ \href{https://orcid.org/0000-0000-0000-0000}{\includegraphics[scale=0.06]{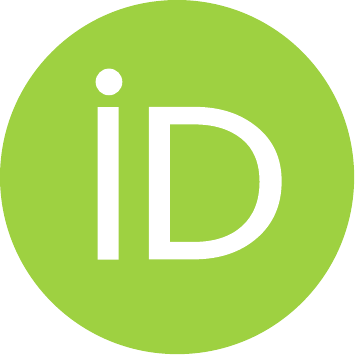}\hspace{1mm}Hakim Benkirane}\thanks{Corresponding author} \\
	Laboratory of Mathematics and Computer Science\\
	CentraleSupélec, Université Paris-Saclay\\
	GIf-sur-Yvettes, 91170 \\
	\texttt{hakim.benkirane@centralesupelec.fr} \\
	\And
	\href{https://orcid.org/0000-0000-0000-0000}{\includegraphics[scale=0.06]{orcid.pdf}\hspace{1mm}Yoann Pradat} \\
	Laboratory of Mathematics and Computer Science\\
	CentraleSupélec, Université Paris-Saclay\\
	Gif-sur-Yvettes, 91170\\
	\texttt{yoann.pradat@centralesupelec.fr} \\
	\And
	\href{https://orcid.org/0000-0002-6963-2968}{\includegraphics[scale=0.06]{orcid.pdf}\hspace{1mm}Stefan Michiels} \\
	CESP-Oncostat Team\\
	GustaveRoussy\\
	Villejuif, 91170\\
	\texttt{stefan.michiels@gustaveroussy.fr} \\
	\And
	\href{https://orcid.org/0000-0001-7679-6197}{\includegraphics[scale=0.06]{orcid.pdf}\hspace{1mm}Paul-Henry Cournède}\thanks{Corresponding author} \\
	Laboratory of Mathematics and Computer Science\\
	CentraleSupélec, Université Paris-Saclay\\
	Gif-sur-Yvettes, 91170\\
	\texttt{paul-henry.cournede@centralesupelec.fr} \\
}

\renewcommand{\shorttitle}{CustOmics}

\hypersetup{
pdftitle={A template for the arxiv style},
pdfsubject={q-bio.NC, q-bio.QM},
pdfauthor={David S.~Hippocampus, Elias D.~Striatum},
pdfkeywords={First keyword, Second keyword, More},
}

\begin{document}
\maketitle

\begin{abstract}
Recent advances in high-throughput sequencing technologies have enabled the extraction of multiple features that depict patient samples at diverse and complementary molecular levels. The generation of such data has led to new challenges in computational biology regarding the integration of high-dimensional and heterogeneous datasets that capture the interrelationships between multiple genes and their functions. 
Thanks to their versatility and ability to learn synthetic latent representations of complex data, deep learning methods offer promising perspectives for integrating multi-omics data. These methods have led to the conception of many original architectures that are primarily based on autoencoder models. However, due to the difficulty of the task, the integration strategy is fundamental to take full advantage of the sources' particularities without losing the global trends. This paper presents a novel strategy to build a customizable autoencoder model that adapts to the dataset used in the case of high-dimensional multi-source integration. We will assess the impact of integration strategies on the latent representation and combine the best strategies to propose a new method, CustOmics (\url{https://github.com/HakimBenkirane/CustOmics}). We focus here on the integration of data from multiple omics sources and demonstrate the performance of the proposed method on test cases for several tasks such as classification and survival analysis.
\end{abstract}

% keywords can be removed

\section{Introduction}
With the advent of high-throughput technologies, multiple omics data are more and more available to characterize the molecular portraits of patients, notably in oncology. Analyzing this mix of data sources and leveraging their information to improve our understanding of biological phenomena remain challenging. 

Several studies and projects have made available cohorts of data that are characterized by several molecular sources. The Cancer Genome Atlas (TCGA, \url{https://portal.gdc.cancer.gov/}) has profiled thousands of tumor samples for multiple molecular assays and made available several types of data such as genome sequencing, RNA sequencing, DNA methylation, proteomics, etc. Integrating such diversified data is crucial to reduce the uncertainty over a biological process due to various experimental setups and to unravel interactions that a single source cannot extract.

This high diversity of data comes with two significant challenges. The first one is linked to the high-dimensionality of the data. Due to the genetic complexity of the human molecular profile, omics data generally suffers from the '\textit{curse of dimensionality}' resulting from the high number of features compared to the smaller number of samples \cite{omics_curse_dim}. This high-dimensional space often contains correlated features that result in high redundancy, reducing the prediction performance of algorithms \cite{ML_redundance}. The second issue lies in data heterogeneity: having its origins from different sources and expressing different phenomena in humans' biological system, omics data is very diverse \cite{omics_het}. For example, transcriptomics and proteomics are normalized differently from other omics data and use different scaling before their analysis, leading to different ranges and data distributions. Finally, omics data like metabolomics can also generate sparsity, as some variables can be below the detection limit and hence be assigned null values.

To overcome those challenges and alleviate overfitting in prediction tasks, standard methods either manually select a small subset of molecular features based on domain knowledge \cite{biofilter} or use dimensionality reduction techniques before downstream analysis. However, due to the high heterogeneity of the sources, those methods can overlook some genome-wide hidden patterns that justify the need for more flexible methods.

In the past few years, multi-omics integration has been a very active research subject in health science and precision medicine \citep{Zitnik2019}. Multiple statistical learning methods have been proposed from the desire to investigate various complex molecular systems behind cancer. One of the most well-known methods engraved in the statistical field is the Principal Component Analysis (PCA) (extensively presented, for example, in \cite{pca}). From this standard model, several variations have been explored, which include Multiple Factor Analysis \citep{mfa}, Consensus PCA, and multi-block PCA \citep{multi_block_pca}. All those methods try to extend the Principal Component approach to a multi-source framework by considering each source as a distinct block. Another variation of the PCA is the Non-negative Matrix Factorization (NMF) first introduced in \cite{nmf} which uses the same philosophy as PCA but considers a non-negative constraint instead of an orthogonality one. Additionally, we can consider the joint Dimensionality Reduction methods (jDR) as an extension of factorial methods studied in a comprehensive benchmark in \cite{cantini}. Such methods can be divided into several categories including Bayesian methods \citep{bayesian_clustering} that use assumptions on data distribution and dependencies to build a statistical model, as well as network-based methods that rely on a network representation and can identify modules of the disease-associated mechanisms. The genes are represented by nodes, and edges represent the association between those genes \citep{paradigm, lemon_tree}. Other more specific methods have been explored, mainly for clustering. For example \cite{autosome} that presented AutoSOME which is a framework using the Self-organizing Map approach and couples it with a density equalizer; this equalizer reduces dimension and preserves the local topology of the gene expressions. Another well-known multi-omics clustering method is iCluster \citep{icluster}, which uses the same idea as NMF but without the non-negative constraint and allows more diversity in data distributions.

With the rise of deep learning as the new state-of-the-art approach in many medical applications, several studies have explored the interest of these methods for multi-omics data integration. The primary use of these methods has been with autoencoders \cite{autoencoders}, notably by \textit{Chaudhary et al.} \cite{autoencoders_multi_omics} in the context of survival prediction. The framework OmiVAE \citep{omivae} introduced new improvements for standard representation learning and OmiEmbed \citep{omiembed} for multitask learning. \citep{simidjievski} also proposed multiple deep learning architectures based on variational frameworks for multiple tasks. \textit{Huang et al.} \citep{salmon} developed a neural network architecture to integrate multi-omics data for survival by feeding the eigengene matrix instead of the raw data. Based on these methods, several application studies on specific cancer cases were presented, like \cite{ov_multi_omics} that used the OmiVAE framework to conduct an in-depth study of the Ovarian Cancer's mechanisms. \cite{feature-level} also used standard autoencoders to explore multi-view learning in survival analysis for Breast Cancer. 
%However, even if \cite{feature-level} showed a glimpse of what autoencoders could do in the context of multi-view learning,
To our knowledge, no benchmarking study has explored and compared the different deep learning approaches and strategies for multi-omics data integration in the context of multitask learning.
%, taking into account not only dimensionality reduction but also the diversity in the ways of dealing with the integration of information in the context of multitask learning.

In this work, we discuss different strategies for the integration of high-dimensional multi-source data to learn low-dimensional latent representation from multi-omics datasets. Moreover, we introduce a new architecture called CustOmics that combines the advantages of the different strategies and alleviates some of their limitations. We evaluate the impact of this new method on different test cases: we first apply it to a pan-cancer dataset and then study how it handles smaller datasets by evaluating it on specific cohorts. We also provide a new package to help bioinormaticians and computational biologists build deep learning architectures that allow an easy switch between the multiple strategies to adapt to specific use cases.

\section*{Materials and methods}

\subsection*{Representation Learning for Multi-Omics Integration}

Representation Learning is a field of statistical learning that aims to automatically discover relevant representations of the input data \citep{bengio2013}. In the field of multiple omics data integration, it can help synthesize the heterogeneous distributions into a common space, revealing the interaction between the different sources. 
From a biological point of view, each omics source represents a specific view of the processes involved. Therefore, the main goal behind integration is to uncover a broader view of the phenomenon of interest and to better understand the underlying processes. 

Each patient is characterized by $K$ omics vectors $(\mathbf{x}_k)_{1\leq k K} \in \mathbb{R}^M_k$ with $M_k$ the number of features of the $k$-th source. The models presented in this paper aim at mapping the set of omics vectors to a vector $\mathbf{z} \in \mathbb{R}^m$ with $m \ll \sum_{k=1}^{K}M_k$, the latent representation. The latent features are the components of the vector $\mathbf{z}$. We will use factorial methods as a basis of comparison as they have proved to be the reference method for mapping input data into a latent space of lower dimension, notably for multi-omics integration, as illustrated by the recent review of \textit{Cantini et al.} \cite{cantini}.

For factorial analysis, we consider approaches based on the study of principal components using linear operations through Multiple Factor Analysis (MFA).
Deep learning methods use autoencoder architectures to build the latent representation by jointly training encoder and decoder functions. The types of architecture can be divided into three main categories:
\begin{itemize}
    \item \textbf{Early Integration (EI):} Refers to the set of methods that aim at merging the different sources before the dimensionality reduction. Instead of giving as model inputs a set of separate vectors for each source, we first concatenate the vectors of all the sources. It is characterized by its simplicity, but finds its limitations when some sources bear more significant signals than the others, which makes it difficult for the model to learn interactions between sources.
    \item \textbf{Joint Integration (JI):} In this setting, sub-representations are created inside the same model for each source before learning the output. This is the most promising strategy theoretically and is the most popular in the literature. However, it is challenged by the sources' heterogeneity, as they do not necessarily follow the same learning dynamics and may need different approaches with specific losses.
    \item \textbf{Late Integration (LI):} Consists in learning an output for each source separately, with its own representation model. It can adapt well to the specificities of each source, but does not retrieve any cross-modality interaction.
\end{itemize}

\begin{figure*}[!t]
\includegraphics[width=18cm, height=7cm]{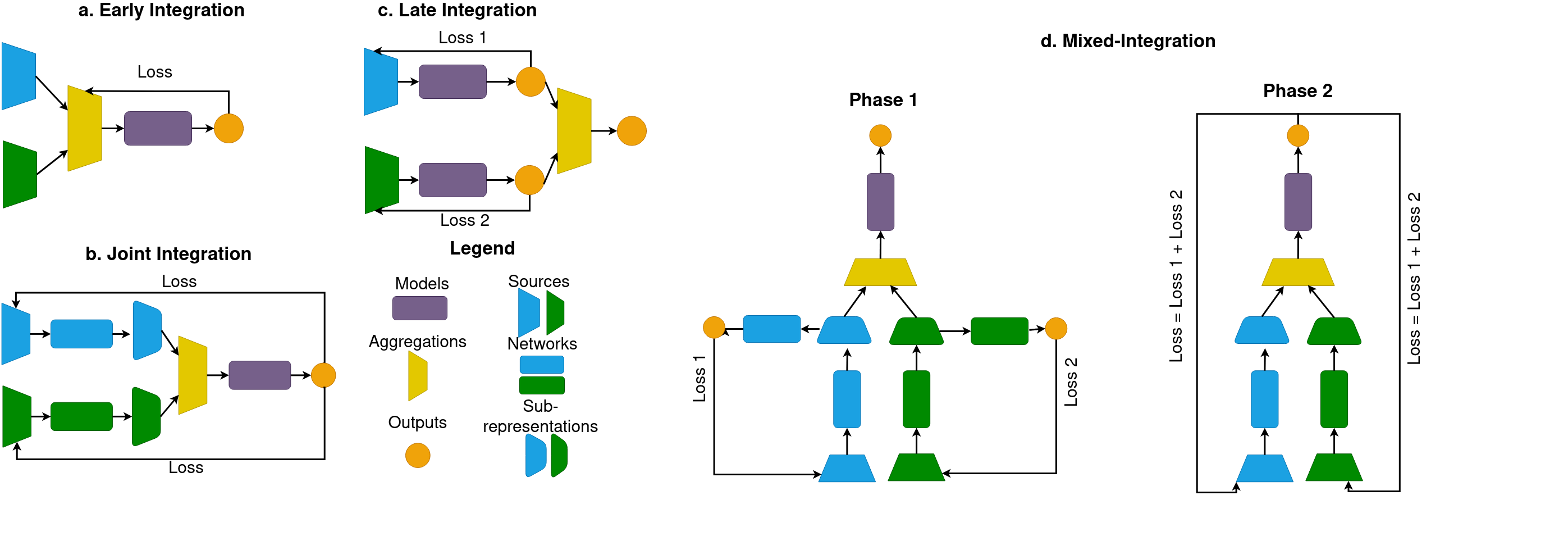}
\caption{\textbf{a. Early Integration:} Sources are concatenated before being fed to a single model. \textbf{b. Joint Integration:} Sub-representations of each source are learned jointly before inputting into the model. \textbf{c. Late Integration:} Each of the sources outputs its prediction using its independent model, the predictions are then aggregated. \textbf{d. Mixed Integration:} Representation of the mixed integration approach. In phase 1, a specific model is trained for each source independently and embeds a sub-representation adapted to the source's specificities. In phase 2, the specific models are trained jointly to create the final output.}
\label{fig:overview}
\end{figure*}

To alleviate the issues brought by those different integration methods, we propose a hybrid strategy, named mixed-integration, that comes halfway between the joint and late integration. It consists of two learning phases: the first phase independently trains the network of each source with an adapted loss to create sub-representations. Those specific models will then be jointly trained in a second phase to build a global output representation. That way, we can adapt our training to each source that may have different learning dynamics in the first phase, while still learing cross-modality interactions in the second phase.

\subsection*{Variational Autoencoders}

A Variational Autoencoder (VAE) is a deep generative model which can learn meaningful data representations from high-dimensional input data. It is an extension of standard autoencoders in which the encoder encodes the input as a distribution over a latent space instead of a single point.

In this case, the encoding function $q$ represents a variational distribution (known as encoding distribution) $q_{\phi}(\mathbf{z}|\mathbf{x})$ in which $\phi$ is the parameter to estimate 
and the decoding function $p$ represents the posterior $p_{\theta}(\mathbf{x}|\mathbf{z})$.

The particularity of a VAE is its ability to encode a distribution. After the encoding phase, there is a sampling phase in which we sample points from the distribution $q_{\phi}(\mathbf{z}|\mathbf{x})$.

Traditionally, the distributions in the VAE architecture are supposed Gaussian: the encoder function will learn the two parameter vectors $\mathbf{\mu}$ and $\mathbf{\sigma}$ (respectively the means and standard deviations) that are used to generate samples in $q_{\phi}(\mathbf{z}|\mathbf{x})$ using the reparametrization trick $\mathbf{z} = \mathbf{\mu} + \mathbf{\sigma}\odot\mathbf{\epsilon}$ where $\mathbf{\epsilon} \sim \mathcal{N}(0,I)$ and $\odot$ denotes the element-wise product.

The loss function for this architecture can be written as the sum of two distinct losses. First, a reconstruction loss that focuses on the autoencoder's ability to reconstruct the data: 

\begin{equation}
    \mathcal{L}_{recon} = \mathbb{E}_{q_{\phi}(\mathbf{z}|\mathbf{x})}[p_{\theta}(\mathbf{x}|\mathbf{z})]
\end{equation}

It can be interpreted as the conditional entropy of $x$ over $z$ which quantifies the amount of uncertainty that one has over the joint distribution $(x, z)$, knowing $z$. In a more practical way, it is related to the quantity of information of $x$ that is retained by $z$ (thus qualifying a reconstruction potential).

Second, a regularization loss aims at getting the encoding distribution as close as possible to the theoretical distribution of the latent vector. Traditionally, a Kullback-Leibler divergence is used: $\displaystyle \mathcal{L}_{reg} = D_{KL}(q_{\phi}(\mathbf{z}|\mathbf{x})||p_{\theta}(\mathbf{z}))$.

% \begin{equation}
%     \mathcal{L}_{reg} = D_{KL}(q_{\phi}(\mathbf{z}|\mathbf{x})||p_{\theta}(\mathbf{z}))
% \end{equation}

The total loss will benefit from the framework introduced in \cite{beta_vae}, $\displaystyle \mathcal{L} = \mathcal{L}_{recon} + \beta\mathcal{L}_{reg}$, adding a weight on the regularization term to balance between the two parts of the loss function, with a hyperparameter $\beta$ to optimize.

\begin{figure*}[h!]
\includegraphics[width=15cm, height=10cm]{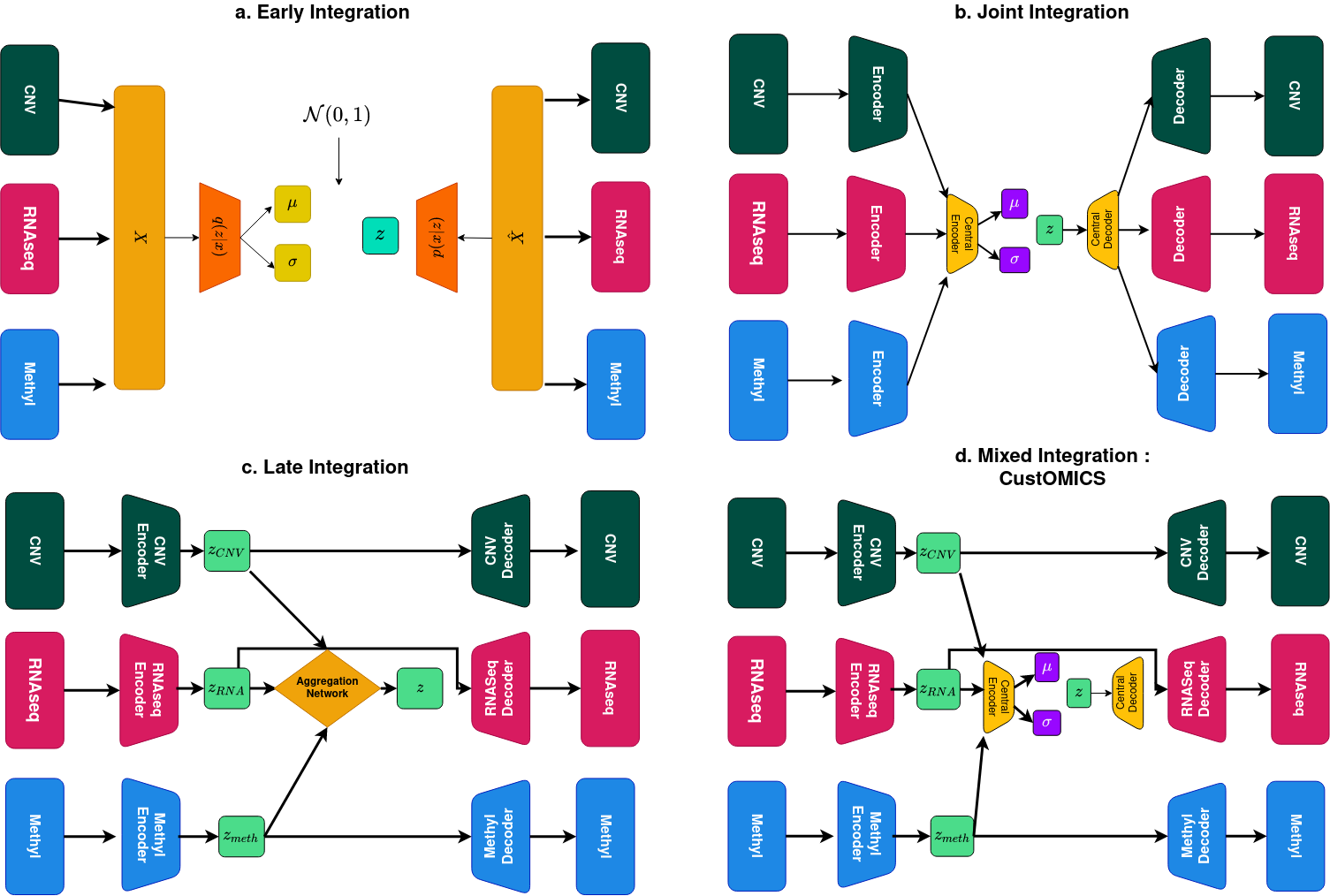}
\caption{\textbf{a. Early Integration VAE:} Variational Autoencoder architecture with early integration strategy. \textbf{b. Joint Integration VAE:} Variational Autoencoder architecture with joint integration strategy. \textbf{c. Late Integration VAE:} Variational Autoencoder architecture with late integration strategy.\textbf{d. Mixed-Integration/CustOmics:} This is a hierarchical architecture composed of specific per-source autoencoders that converges into a central variational autoencoder.}
\label{fig:vae}
\end{figure*}

\subsection*{Early Integration}

An early integration autoencoder is an autoencoder applied to the concatenated data from various sources. This type of architecture was reported in \cite{gautier} as the SDAE (Standard Deep Autoencoder). The drawback of such an integration is that the resulting concatenation of every dataset results in a noisier and more complex high-dimensional matrix that makes the training more difficult. Moreover, this type of integration does not consider the specific data distribution of each source separately and thus leads the model to learn irrelevant patterns that can, in the worst case, mask the essential signals.
For those reasons, even though this strategy has been introduced in the benchmark, it is not recommended in a real setting for complex multi-source datasets such as omics data.
For comparison purposes, we adapt the SDAE architecture to a Variational Autoencoder instead of a standard autoencoder as described in Figure \ref{fig:vae}

\subsection*{Joint Integration}

This is a standard way to approach multi-source integration using Variational Autoencoders. This model has been used in multiple designs in literature (\cite{omivae}; \cite{omiembed}; \cite{simidjievski}) and is probably the most popular as it is relatively straightforward.
The idea is to reduce through a series of the hidden layer the dimension of each source separately, then concatenate them in a central layer before sampling.
In this architecture, we would learn the mean and standard deviation of the concatenated data such that:

\begin{itemize}
    \item $\mathbf{z} = \mathcal{N}(\mu, \sigma^2I)$
    \item $\mu = q_{\mu}(T(q^{(1)}(\mathbf{x}_1; \theta_{e,1}),..., q^{(K)}(\mathbf{x}^{(K)}; \theta_{e,K})))$
    \item $\sigma = q_{\sigma}(T(q^{(1)}(\mathbf{x}^{(1)}; \theta_{e,1}),..., q^{(K)}(\mathbf{x}^{(K)}; \theta_{e,K})))$
\end{itemize}

\subsection*{Late Integration}

A late integration approach for variational autoencoders is a variation of the architecture proposed in \cite{simidjievski} with the Hierarchical VAE architecture. The idea is to use variational inference on each layer separately and input the resulting representations in an aggregation network learned independently of the other VAEs, which is the main difference from the joint integration approach. This architecture is described in Figure \ref{fig:vae}.

\subsection*{Mixed Integration / CustOmics}

This integration strategy will be the foundation for building the CustOmics architecture. The proposed method is a hierarchical mixed-integration that consists of an autoencoder for each source that creates a sub-representation that will then be fed to a central variational autoencoder.
This new integration strategy benefits from two training phases. The first phase will act as a normalization process: each source will train separately to learn a more compact representation that synthesizes its information with less noise. This will help the integration as we will lose all imbalance issues between the sources and avoid loss of focus when a source has an inferior dimensionality or weaker signal than the others.
The second phase will constitute a simple joint integration between the learned sub-representations, while still training all the encoders to fine-tune those representations as some signals are enhanced in the presence of other sources.
Regarding the regularization loss for the central layer, the KL divergence can be an obstacle to generalization. As stated in \cite{infovae}, the KL divergence suffers from various problems. The first is that the model can fail to learn a meaningful representation of the input. Indeed, the KL divergence can sometimes be too restrictive and naturally tends to make the latent code a random sample from $p_{\theta}(Z)$. The second is that the KL divergence can make the model overfit and learn a latent code with variance tending to infinity.

So instead, we will use the Maximum Mean Discrepancy (MMD) to assess the distance between the distributions. This distance stands on the foundation that two distributions are identical if and only if their moments are the same. Let $p,q$ be two distributions, the MMD distance is given as follows:

\begin{equation}
    MMD(p(x)||q(x)) = \mathbb{E}_{p(x), p(x')}(\kappa(x, x'))+ \mathbb{E}_{q(x), q(x')}(\kappa(x, x'))
    -2\mathbb{E}_{p(x), q(x')}(\kappa(x, x'))
\end{equation}

where $\kappa$ is a kernel function. We will choose a Gaussian kernel $\kappa(x, x') = e^{\frac{-||x - x'||^2}{2\sigma^2}}$.

Whereas all the models mentioned previously build the latent representation in an unsupervised way, we also create latent features adapted to specific tasks like classification or survival. This idea has been used multiple times in the literature, for example in \cite{omiembed}.

The solution relies on adding a task-related loss to the autoencoder objective function. Therefore, we denote by $\mathcal{L}_{task}$ the loss such that $\displaystyle \mathcal{L}_{tot} = \mathcal{L}_{AE} + \alpha\mathcal{L}_{task}$. 

For the classification task, we use a categorical cross-entropy loss defined by $\displaystyle \mathcal{L}_{class} = \sum_{i}y_ilog(\hat{y}_i)$, where $y_i$ is the ground truth for the $i^{th}$ sample, and $\hat{y}_i$ its estimation with the downstream model.

For the survival task, we use the DeepSurv loss function. This nonlinear proportional hazard model is introduced in \cite{deepsurv}. The model is built by using the negative partial log-likelihood formula that translates in our case into:
\begin{equation}
    L(\theta) = -\sum_{i:E_i = 1}(\hat{\mu}(t, x_i; \theta) - log \sum_{j \in \mathcal{R}(T_i)} e^{\hat{\mu}(x_i; \theta)})
\end{equation}
where $\hat{\mu}(x; \theta)$ is the risk function estimated by the output layer of the network, $\mathcal{R}(t)$ is the risk set, that is the set of patients still at risk of failure after time $t$.

Additionally, this loss will be assigned to each omic-specific network in the first training phase to create adequate sub-representations before the joint integration phase.

\subsection*{Interpretability}

To build a more interpretable architecture, we adapt the method introduced in \cite{XOmiVAE2021} to compute SHAP values \citep{shap} for deep variational autoencoders. Whereas this method has only been conceived for single source inputs, we expand it to the multi-source setting of CustOmics by adapting it to any deep autoencoder and applying it to each source autoencoder thanks to the dual-phase approach characterizing the mixed-integration strategy.
After training the CustOmics network, we pass each intermediate autoencoder to a DeepSHAP explainer that was modified to make sure that it could take either each component of the latent vector or the prediction output for the contribution analysis. For each prediction, SHAP values corresponding to each gene or latent dimension are computed to estimate the overall contribution by averaging them over a group of samples with similar features.
When combined with the mixed-integration approach through CustOmics, we can compute for each source two types of features importance. During the first training phase, we compute the features importances of each source separately to assess the importance of genes when only considering this source. During the second training phase, the features' importance can change  as it will then consider other sources, showing how some genes can have increasing or decreasing importance when introducing complementary views of the biological system.

\subsection*{Test Cases and Datasets}

In this study, we use datasets extracted from the Genomic Data Commons (GDC) pan-cancer multi-omics study \cite{pancan}. It is one of the most comprehensive datasets for multi-omics analysis, with high-dimensional omics data and corresponding phenotype data from The Cancer Genome Atlas (TCGA).
In our experiments, we use three types of omics data, Copy Number Variations (CNV), RNA-Seq gene expressions, and DNA methylation. The CNV dataset comprises Gistic2 measurements on a total of 19,729 genes. The RNA-Seq expression dataset profile comprises around 60,484 identifiers referring to corresponding exons and measuring log2 transformed Fragments Per Kilobase of
transcript per Million mapped reads (FPKM). Finally, the DNA methylation dataset was produced using the Infinium HumanMethylation450 BeadChip (450K) arrays with 485,578 probes in which beta values of probes indicate the methylation ratio of corresponding CpG sites.
Moreover, we also evaluate our method on six smaller cohorts from TCGA: Bladder Urothelial Carcinoma (BLCA, n=437), Breast Invasive Carcinoma  (BRCA, n=1022), Lung Adenocarcinoma (LUAD, n=498), Glioblastoma \& Lower Grade Glioma  (GBMLGG, n=515) and Uterine Corpus Endometrial Carcinoma  (UCEC, n=538).
We will perform in this study 4 different evaluations on 4 test cases for clasification and survival. The first task in our set of experiments is classifying the different tumor types in the pan-cancer study. The classification performance was measured using five metrics: Accuracy, Macro-averaged F1 score, precision, recall, and ROC-AUC \cite{roc-auc}. We also perform a second classification task to validate our findings on a smaller dataset and test the robustness of our method. This task aims at performing a tumor subtype classification based on the PAM50 classification (IlluminlA, ILluminalB, Basal, and HER2), we use the same setup as the pan-cancer case. The third test case will be a survival study of the Pancancer dataset. Finally, the fourth test case will be about evaluating the survival performances on the 5 datasets presented earlier. For all those test cases, We compare the CustOmics model to several reference methods for multi-omics integration: first with a combination of dimensionality reduction methods, Multiple Factor Analysis (MFA), Uniform Manifold Approximation and Projection (UMAP) and non-negative matrix factorization (NMF), and also with different deep learning methods corresponding to the various strategies described in the Methods section.

\subsection*{Data Preprocessing}

The data required some preprocessing before analysis.
\begin{itemize}
    \item For RNA gene expression profiles, 594 exons located on the Y chromosome were removed, along with 1,904 ones with zero expression and 248 with missing values.
    \item For DNA methylation data, the same strategy as with gene expression profiles was used, in addition to removing probes that cannot be mapped to the human reference genome. It leaves us with 438,831 CpG sites.
\end{itemize}

Afterward, we intersected each combination of omics data in order to retrieve the maximum number of samples for each test case. We then identified and removed features with missing/consistently zero/NA values for other omic files. Finally, for non-normalized datasets such as CNVs and RNA-Seq data, we applied a min-max normalization to ensure that each omic source was scaled identically and thus would have the same importance during integration.

\subsection*{Implementation Details}

The CustOmics framework is based on the Pytorch deep-learning library. It can be applied to any combination of high-dimensional datasets and incorporates different integration strategies depending on the type of data and task to perform. As done in \textit{Zhang et al.} \cite{omiembed}, DNA methylation data can be divided into 23 separate blocks, each feeding a hidden layer corresponding to a chromosome to avoid overfitting and save GPU memory.

The whole architecture is built using fully-connected blocks. We use a batch normalization technique in each layer composing the neural network to address the internal covariate shift problem\cite{batch_norm}. Also, to avoid overfitting problems, we use dropout \cite{dropout}; its rate is considered a hyperparameter.

The input dataset was randomly split into training, validation, and testing set (60-20-20\%) by using a stratified 5-fold cross-validation so that the proportion of samples in each tumor type between the different sets is preserved in all the folds. We perform Bayesian optimization \cite{bayesian_optim} using the validation set to find our model's best possible combination of hyperparameters.

All models were trained using an Nvidia Tesla V100S with 32 GB memory.

% Results and Discussion can be combined.
\section*{Results}

\subsection*{Classification Results}
\label{sec:classifresults}

 We first perform the classification task on the pan-cancer dataset. Each architecture is coupled with an artificial neural network classifier composed of two hidden layers with 256 and 128 neurons.

\begin{figure*}[!t]
\includegraphics[width=17cm, height=18cm]{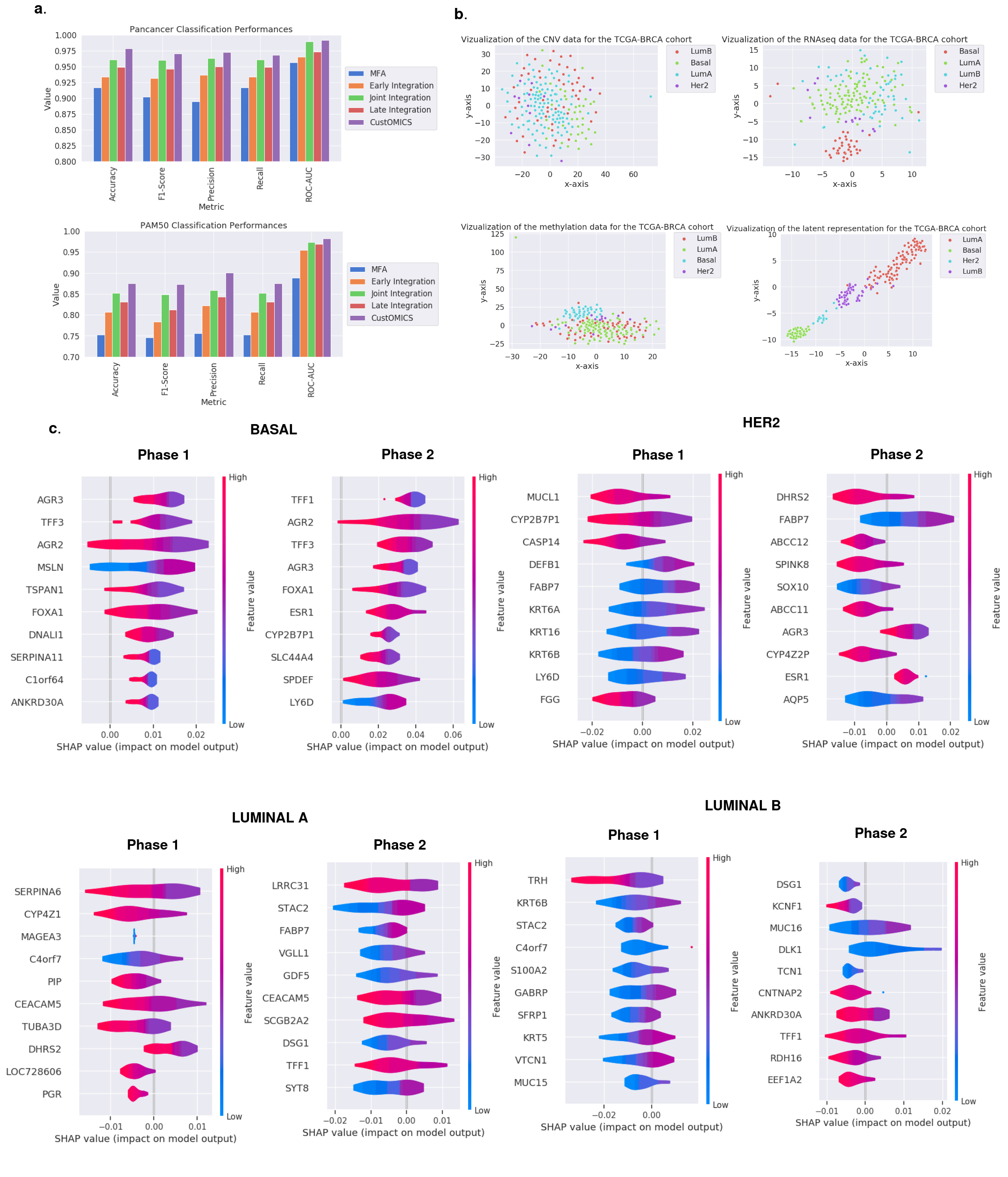}
\caption{\textbf{a. pan-cancer and PAM50 classification results:} Overall classification results for the pan-cancer tumor classification test case and the PAM50 subtype classification for breast cancer. \textbf{b. T-SNE vizualization} for each omic source separately, along with the latent representation constructed by CustOmics. We see that the constructed layer representation succeeds at separating the data into 4 distinct clusters that we couldn't distinguish with each omic source alone. \textbf{c. PAM50 gene importance:} Computed SHAP values on the RNA-Seq data of the most relevant genes responsible for the discrimination between each subtype against the others using CustOmics for both integration phases.}
\label{fig:classif-results}
\end{figure*}

\begin{table}[h!]
\caption{The classification performance for the pan-cancer dataset is evaluated with 5 standard metrics for UMAP, NMF, MFA, and deep-learning methods. We evaluate the performances on the final predicted output of the downstream classifier.}
\begin{tabular}{|c|c|c|c|c|c|}
\hline
Model & Accuracy & F1-score & Precision & Recall & ROC-AUC \\
\hline
UMAP & $0.7598 \pm 0.0036$ & $0.7149 \pm 0.0029$ & $0.7200 \pm 0.0031$ & $0.7598 \pm 0.0032$ & $0.8740 \pm 0.0012$ \\
NMF & $0.8599 \pm 0.0017$ & $0.8406 \pm 0.0013$ & $0.8460 \pm 0.0018$ & $0.8599 \pm 0.0021$ & $0.9266 \pm 0.0019$ \\
MFA & $0.9167 \pm 0.0012$ & $0.9025 \pm 0.0014$ & $0.8945 \pm 0.0008$ & $0.9167 \pm 0.0013$ & $0.9565 \pm 0.0003$ \\
\hline
Early Int. VAE & $0.9337 \pm 0.0079$ & $0.9314 \pm 0.0086$ & $0.9367 \pm 0.0067$ & $0.9337 \pm 0.0079$ & $0.9655 \pm 0.0041$ \\
Joint Int. VAE & $0.9610 \pm 0.0032$ & $0.9600 \pm 0.0032$ & $0.9631 \pm 0.0043$ & $0.9610 \pm 0.0032$ & $0.9898 \pm 0.0005$ \\
Late Int. VAE & $0.9492 \pm 0.0115$ & $0.9464 \pm 0.0111$ & $0.9498 \pm 0.0079$ & $0.9492 \pm 0.0115$ & $0.9737 \pm 0.0060$ \\
\textbf{CustOmics} & $\mathbf{0.9788 \pm 0.0025}$ & $\mathbf{0.9705 \pm 0.0033}$ & $\mathbf{0.9728 \pm 0.0041}$ & $\mathbf{0.9685 \pm 0.0034}$ & $\mathbf{0.9918 \pm 0.0001}$ \\
\hline
\end{tabular}
  \label{tab:pancan-results}
\end{table}

Figure \ref{fig:classif-results} and table \ref{tab:pancan-results} show the overall classification results (More details can be found in Table S1). Among the factorial methods, the MFA achieved the best results, so we coupled this method with the same ANN classifier used with the deep-learning representation methods as a basis for comparison. However, it does not perform as well as most deep-learning methods. It is because MFA cannot uncover nonlinear relationships between different sources, unlike deep-learning architectures. Moreover, as the MFA is an early integration, it suffers from the problems mentioned above when we introduced this strategy.

As hinted earlier, we can see that for the deep learning strategies, early integration is behind the others in terms of performance. It can be explained by the fact that RNA-Seq data hold more signals when determining tumor types or subtypes. Thus, concatenating the sources before feeding them to the VAE overshadows the other sources, and the learned representation depends mostly on RNA-Seq data without leveraging the other modalities. It is illustrated in Table S2, showing that the classification results using RNA-Seq data only are very close to those obtained with early integration, indicating that the model may overlook the interactions between sources.
Late integration is not optimal either, since interactions between sources are not properly learned.
Joint integration performs well in most cases, but we see that the best results displayed in Table S2 are achieved by the combination of only two sources, RNA-Seq and Methylation data, as it seems that CNV data only adds noise to the latent representation, meaning that its information is not handled well with this strategy.

\begin{table}[h!]
\centering
\caption{Classification performance for PAM50 classification on the TCGA-BRCA dataset, with 5 standard metrics. We compare machine learning methods like UMAP, NMF, and MFA with deep-learning methods. We evaluate the performances on the final predicted output of the downstream classifier.}
\begin{tabular}{|c|c|c|c|c|c|}
\hline
Model & Accuracy & F1-score & Precision & Recall & ROC-AUC \\
\hline
UMAP & $0.6815 \pm 0.0152$ & $0.6612 \pm 0.0140$ & $0.6637 \pm 0.0157$ & $0.6815 \pm 0.0151$ & $0.8482 \pm 0.0055$ \\
NMF & $0.7025 \pm 0.0132$ & $0.6955 \pm 0.0135$ & $0.7032 \pm 0.0110$ & $0.7025 \pm 0.0131$ & $0.8576 \pm 0.0119$ \\
MFA & $0.7532 \pm 0.0164$ & $0.7460 \pm 0.0160$ & $0.7562 \pm 0.0162$ & $0.7531 \pm 0.0165$ & $0.8884 \pm 0.0037$ \\
\hline
Early Int. VAE & $0.8063 \pm 0.0152$ & $0.7840 \pm 0.0167$ & $0.8228 \pm 0.0167$ & $0.8063 \pm 0.0150$ & $0.9552 \pm 0.0077$ \\
Joint Int. VAE & $0.8518 \pm 0.0184$ & $0.8488 \pm 0.0189$ & $0.8589 \pm 0.0161$ & $0.8518 \pm 0.0151$ & $0.9734 \pm 0.0035$ \\
Late Int. VAE & $0.8312 \pm 0.0174$ & $0.8124 \pm 0.0201$ & $0.8429 \pm 0.0215$ & $0.8312 \pm 0.0176$ & $0.9689 \pm 0.0066$ \\
\textbf{CustOmics} & $\mathbf{0.8758 \pm 0.0162}$ & $\mathbf{0.8728 \pm 0.0141}$ & $\mathbf{0.9012 \pm 0.0137}$ & $\mathbf{0.8758 \pm 0.0130}$ & $\mathbf{0.9828 \pm 0.0022}$ \\
\hline
\end{tabular}
  \label{tab:brca-results}
\end{table}

These results confirm the interest of the CustOmics architecture, as it gives the best performances for all the test cases. Not only does it perform better than the others, but it also takes advantage of the complementarity and interactions between sources: As shown in Table S1, all sources bring an additional amount of information. Figure \ref{fig:classif-results} gives a visualization of the different sources: Even though the initial sources are quite entangled, the CustOmics latent representation succeeds in separating the clusters using the mutual information between modalities, resulting in better classification performances.
We also use the interpretability property of CustOmics introduced in the Methods section to highlight the most relevant features for the discrimination between PAM50 subtypes by computing their respective SHAP values for each source. We do it for both phases: in phase 1, we retrieve the relevant genes considered when using a single omic source, whereas, in phase 2, we investigate how the addition of other sources' signals changes the genes' importance. Figure \ref{fig:classif-results} references the results of such explanations on RNA-Seq data, Figure S1 and S2 show the results for CNV and methylation data. We witness some well-referenced biomarkers for breast cancer like TFF1 \cite{tff1_biomarker}, suggesting that our method can retrieve relevant biological information.

\subsection*{Survival Analysis}

The second task in this study is survival analysis. The goal is to predict the risk score associated with each patient from the corresponding high-dimensional omics data.
Two standard metrics evaluate the performance of this downstream task: The c-index, which is a generalization of the AUC metric for censored data \citep{harrell}, and the Integrated Brier Score \citep{ibs}.

\begin{figure*}[!t]
\centering
\includegraphics[width=19cm, height=13cm]{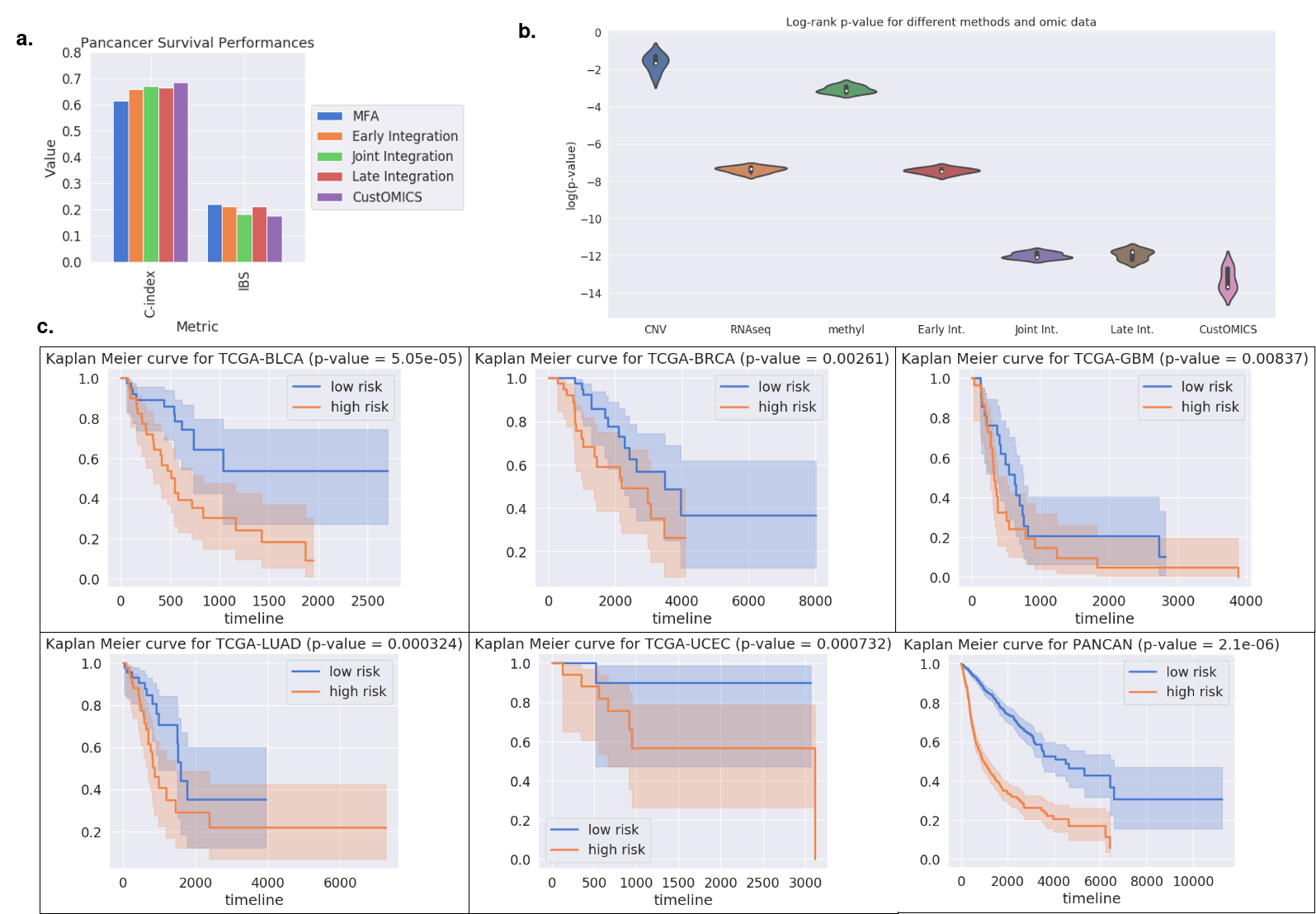}
\caption{\textbf{a. Survival Analysis Performances:} We evaluate the performances of the survival model for the pan-cancer dataset using both C-index and the Integrated Brier Score (IBS). Here again, our model outperforms the other integration strategies on both metrics.
\textbf{b. Log-rank test:} We compute the p-value associated to the log-rank test between high and low risk groups for every integration strategy on a validation set for the pan-cancer survival test case, and we compare it to mono-omic survival predictions. \textbf{c. Kaplan Meier Curves :} We draw the Kaplan Meier curves and display the p-value associated to the log-rank test as computed previously for each dataset using the predicted hazard from the CustOmics model and stratify the population into high and low risk on the test set for the predicted hazard ratio. This figure shows that our method succeeds in stratifying the patients into risk subgroups.}
\label{fig:survival-results}
\end{figure*}

Figure \ref{fig:survival-results} and table \ref{tab:survival-perf} show the results for the different methods (more details can be found in Table S3) for the survival task. The same observations as for the classification task can be done regarding the differences between integration strategies. Here again, we also evaluated the performance of the CustOmics method for each combination of omics sources (Table S3).

\begin{table}[h!]
\centering
\caption{The survival analysis performance for the pan-cancer dataset is evaluated with 2 standard metrics, C-index and IBS. We compare classical methods like UMAP, NMF, and MFA with deep-learning methods and evaluate the performances on the final predicted output of the downstream survival network.}
\begin{tabular}{||c||c||c||}
\hline
Model & C-index & IBS \\
\hline
UMAP & $0.5948 \pm 0.0231$ & $0.2486 \pm 0.0327$ \\
NMF & $0.6012 \pm 0.0204$ & $0.2207 \pm 0.0264$ \\
MFA & $0.6127 \pm 0.0164$ & $0.2192 \pm 0.0203$\\
\hline
Early Int. VAE & $0.6578 \pm 0.0103$ & $0.2106 \pm 0.0117$ \\
Joint Int. VAE & $0.6709 \pm 0.0041$ & $0.1802 \pm 0.0072$ \\
Late Int. VAE & $0.6629 \pm 0.0086$ & $0.2112 \pm 0.0088$ \\
\textbf{CustOmics} & $\mathbf{0.6841 \pm 0.0033}$ & $\mathbf{0.1745 \pm 0.0052}$ \\
\hline
\end{tabular}
  \label{tab:survival-perf}
\end{table}

The last task consists in evaluating the model performances for survival analysis for several specific cancer types of the TCGA datasets described in the dataset section. The objective is to evaluate the robustness of the models when dealing with smaller datasets.

\begin{table}[t!]
\caption{Performances of state-of-the-art integration methods for survival analysis, using concordance index on 5 TCGA cohorts: Bladder Urothelial Carcinoma (BLCA), Breast Invasive Carcinoma (BRCA), Glioblastoma \& Lower Grade Glioma (GBMLGG), Lung Adenocarcinoma (LUAD) and Ulterine Corpus Endometrial Carcinoma (UCEC). }\label{tab:results}
\centering
\scalebox{1.0}{
\begin{tabular}{|l|l|l|l|l|l|l|}
\hline
Model &  BLCA & BRCA & GBMLGG & LUAD & UCEC & Overall\\
\hline
UMAP & $0.527 \pm 0.048$ & $0.524 \pm 0.039$ & $0.557 \pm 0.028$ & $0.530 \pm 0.017$ & $0.543 \pm 0.025$ & $0.536$\\
NMF & $0.553 \pm 0.060$ & $0.560 \pm 0.036$ & $0.584 \pm 0.067$ & $0.589 \pm 0.052$ & $0.570 \pm 0.068$ & $0.571$\\
MFA & $0.591 \pm 0.052$ & $0.599 \pm 0.043$ & $0.605 \pm 0.036$ & $0.597 \pm 0.058$ & $0.586 \pm 0.020$ & $0.596$\\
\hline
Early Int. VAE & $0.603 \pm 0.054$ & $0.618 \pm 0.039$ & $0.628 \pm 0.056$ & $0.612 \pm 0.041$ & $0.609 \pm 0.032$ & $0.614$\\
Joint Int. VAE & $0.616 \pm 0.072$ & $0.627 \pm 0.030$ & $0.635 \pm 0.020$ & $0.608 \pm 0.038$ & $0.630 \pm 0.021$ & $0.624$\\
Late Int. VAE & $0.610 \pm 0.055$ & $0.620 \pm 0.057$ & $0.621 \pm 0.067$ & $0.595 \pm 0.012$ & $0.627 \pm 0.022$ & $0.615$\\
CustOmics & $\mathbf{0.637 \pm 0.050}$ & $\mathbf{0.633 \pm 0.018}$ & $\mathbf{0.642 \pm 0.028}$ & $\mathbf{0.625 \pm 0.037}$ & $\mathbf{0.667 \pm 0.022}$ & $\mathbf{0.640}$\\
\hline
\end{tabular}}
\end{table}

Finally, we perform a more thorough analysis on the survival results that we display in Figure \ref{fig:survival-results}. We leave out 20\% of our datasets for validation purpose, and we perform 5-fold cross-validation on the remaining 80\% to compute the p-values associated to the log-rank test for different combinations of the Pan cancer test case. We show the ability of CustOmics to stratify the patients into distinct risk groups using the predicted hazard ratio. This stratification ability was also measured quantitatively using the p-value associated to the log-rank test between the different categories. Even though the comparison between joint, late and mixed integration is not significant in this case, it is interesting to note that the addition of multiple omics sources has greatly affected the p-value as it was nearly multiplied by a factor $10^{-5}$ for CNV data and $10^{-2}$ for RNA-Seq and methylation data. Those results also show that the early integration is the strategy with minimal enrichment from other sources, which corroborates our previous intuition and is coherent with the results found in the different experiments.

\section*{Discussion}

In this work, we presented a range of integration strategies for multi-source data that can handle both the high dimensionality and the heterogeneity of the data. To take the best of all those strategies, we presented the mixed-integration and the CustOmics framework to alleviate the limitations of the existing methods. This new framework can achieve better latent representations and lead to a more robust and generalizable architecture, as shown with the systematic better results than alternative strategies. Importantly, our method is capable of adapting to each omic source by handling the training independently in a first phase, which solves the issue of unbalanced signals between the sources by standardizing the representations before starting to learn cross-modality interactions.   
Our fusion model demonstrated better performances for both classification and survival outcome prediction across all test cases. It not only achieved great results on pan-cancer data, but it also did on smaller dataset for particular cohorts, showing the robustness of our method in situations with fewer samples.
Interestingly, the performances of CustOmics in comparison to other integration strategies for different combinations of omic sources show that the mixed-integration handles better the fusion between CNV, RNA-Seq and methylation data. This aspect could probably be further enhanced by adding prior knowledge of those sources in the intermediate autoencoders, for example introducing a negative binomial prior in the RNA-Seq autoencoder.
Furthermore, by adapting the SHAP method to our architecture, we were able to highlight genes of importance for specific tasks. However, there is room for improvement from a computational point of view, since the interpretability method is very expensive.

As clinical applications often suffer from a lack of samples to build efficient deep learning models and as all the modalities are not always extracted for all patients, it would be beneficial to study the benefits of per-source transfer learning during phase 1 to pick-up weaker signals and eliminate the noise. In fact, one of the advantages of phase 1 learning the sub-representations independently is that it is not necessary for all patients to have all omic sources available at all times, which helps us take the best advantage of the data at hand.

\section*{Conclusion}

In conclusion, our generic and interpretable multi-source deep learning framework improves on state-of-the-art integration strategies by proposing a hybrid approach that fits well with multi-omics data. The framework is available to use on Github: \url{https://github.com/HakimBenkirane/CustOmics}. Possible steps to refine CustOmics are a more thorough study on the interpretability part of the framework and adding prior knowledge to individual autoencoders to enhance the sub-representations of omic sources during phase 1 of mixed-integration.

\section*{Supplementary Materials}

\subsection*{Classification Results}

\begin{table}[h!]
\centering
\caption{Classification performances for multiple combinations of omics data using CustOMICS on the Pancancer dataset. We can see that the best performances are brought by Transcriptomics data, but the addition of other omics data increases the performances, showing that the integration is relevant.}
\begin{tabular}{||c||c||c||c||c||c||}
\hline
Omics & Accuracy & F1-score & Precision & Recall & ROC-AUC \\
\hline
CNV & $0.47 \pm 0.03$ & $0.47 \pm 0.03$ & $0.47 \pm 0.03$ & $0.48 \pm 0.01$ & $0.75 \pm 0.02$ \\
RNAseq & $0.92 \pm 0.01$ & $0.93 \pm 0.01$ & $0.93 \pm 0.01$ & $0.92 \pm 0.01$ & $0.96 \pm 0.00$ \\
methyl & $0.68 \pm 0.02$ & $0.68 \pm 0.02$ & $0.68 \pm 0.02$ & $0.68 \pm 0.02$ & $0.82 \pm 0.01$ \\
\hline
CNV + RNAseq & $0.93 \pm 0.01$ & $0.92 \pm 0.01$ & $0.92 \pm 0.01$ & $0.92 \pm 0.01$ & $0.96 \pm 0.00$ \\
CNV + methyl & $0.71 \pm 0.02$ & $0.69 \pm 0.02$ & $0.70 \pm 0.02$ & $0.69 \pm 0.02$ & $0.85 \pm 0.01$ \\
RNAseq + methyl & $0.94 \pm 0.01$ & $0.94 \pm 0.01$ & $0.94 \pm 0.01$ & $0.94 \pm 0.0132$ & $0.97 \pm 0.00$ \\
\hline
CNV + RNAseq + methyl& $0.98 \pm 0.01$ & $0.97 \pm 0.01$ & $0.97 \pm 0.01$ & $0.97 \pm 0.01$ & $0.99 \pm 0.00$ \\
\hline
\end{tabular}
  \label{tab:survival-1}
\end{table}

\begin{table}[h!]
\centering
\caption{Classification performances for multiple combinations of omics data using joint integration on the Pancancer dataset.}
\begin{tabular}{||c||c||c||c||c||c||}
\hline
Omics & Accuracy & F1-score & Precision & Recall & ROC-AUC \\
\hline
CNV & $0.47 \pm 0.03$ & $0.47 \pm 0.03$ & $0.47 \pm 0.03$ & $0.48 \pm 0.01$ & $0.75 \pm 0.02$ \\
RNAseq & $0.92 \pm 0.01$ & $0.93 \pm 0.01$ & $0.93 \pm 0.01$ & $0.92 \pm 0.01$ & $0.96 \pm 0.00$ \\
methyl & $0.68 \pm 0.02$ & $0.68 \pm 0.02$ & $0.68 \pm 0.02$ & $0.68 \pm 0.02$ & $0.82 \pm 0.01$ \\
\hline
CNV + RNAseq & $0.90 \pm 0.02$ & $0.89 \pm 0.02$ & $0.90 \pm 0.02$ & $0.88 \pm 0.03$ & $0.93 \pm 0.00$ \\
CNV + methyl & $0.70 \pm 0.02$ & $0.69 \pm 0.02$ & $0.70 \pm 0.02$ & $0.70 \pm 0.02$ & $0.85 \pm 0.01$ \\
RNAseq + methyl & $0.96 \pm 0.01$ & $0.96 \pm 0.01$ & $0.96 \pm 0.01$ & $0.96 \pm 0.0132$ & $0.99 \pm 0.00$ \\
\hline
CNV + RNAseq + methyl& $0.94 \pm 0.01$ & $0.94 \pm 0.01$ & $0.95 \pm 0.01$ & $0.94 \pm 0.01$ & $0.97 \pm 0.00$ \\
\hline
\end{tabular}
  \label{tab:survival-2}
\end{table}

\begin{figure*}[h!]
\centering
\includegraphics[width=16cm, height=13cm]{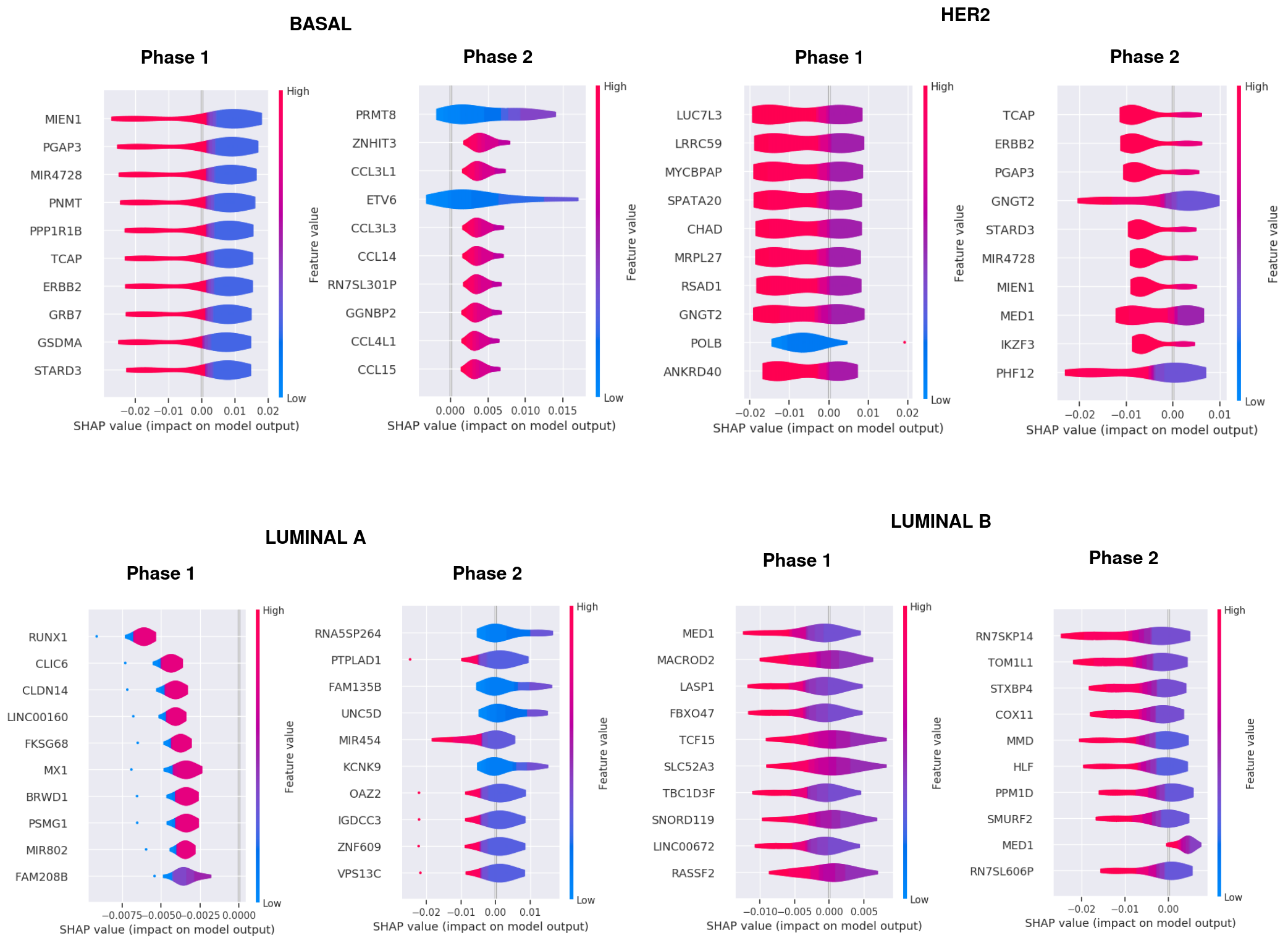}
\caption{PAM50 gene importance: Computed SHAP values on the CNV data of the most relevant genes responsible for the discrimination between each subtype against the others using CustOmics for both integration phases.}
\label{fig:classif-results}
\end{figure*}

\begin{figure*}[h!]
\centering
\includegraphics[width=16cm, height=13cm]{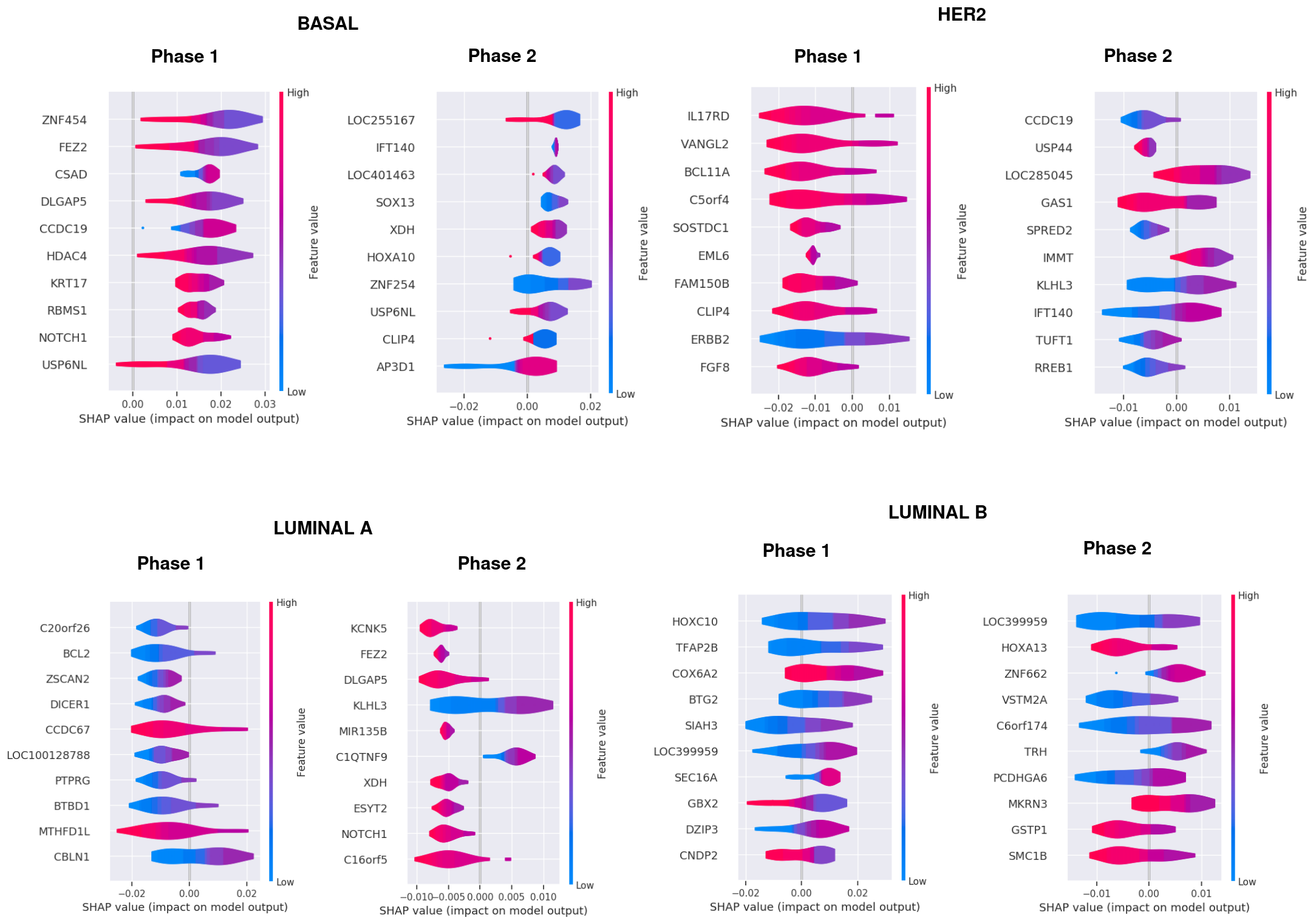}
\caption{PAM50 gene importance: Computed SHAP values on the methylation data of the most relevant genes responsible for the discrimination between each subtype against the others using CustOmics for both integration phases.}
\label{fig:classif-results}
\end{figure*}

\newpage

\subsection*{Survival Results}

\begin{table}[h!]
\centering
\caption{Survival performances for multiple combinations of omics data using CustOMICS on the Pan cancer dataset. We can see that the best performances are brought by Transcriptomics data, but the addition of other omics data increases the performances, showing that the integration is relevant.}
\begin{tabular}{||c||c||c||}
\hline
Omics & Accuracy & F1-score  \\
\hline
CNV & $0.54 \pm 0.05$ & $0.25 \pm 0.06$  \\
RNASeq & $0.63 \pm 0.02$ & $0.20 \pm 0.02$  \\
methyl & $0.59 \pm 0.02$ & $0.23 \pm 0.03$  \\
\hline
CNV + RNAseq & $0.64 \pm 0.02$ & $0.19 \pm 0.02$  \\
CNV + methyl & $0.61 \pm 0.03$ & $0.21 \pm 0.03$  \\
RNAseq + methyl & $0.64 \pm 0.02$ & $0.19 \pm 0.02$  \\
\hline
CNV + RNAseq + methyl& $0.68 \pm 0.01$ & $0.17 \pm 0.01$  \\
\hline
\end{tabular}
  \label{tab:survival}
\end{table}

\bibliographystyle{unsrtnat}
\bibliography{template}  %%% Uncomment this line and comment out the ``thebibliography'' section below to use the external .bib file (using bibtex) .





% \end{thebibliography}

\end{document}